\documentclass[12pt]{article}
\usepackage[english]{babel}
\usepackage{amssymb}
\usepackage{epsfig,color}
\usepackage{chemformula}
\NewCommandCopy{\CH}{\ce}
\let\ce\relax
\usepackage{pstricks,graphicx,epsfig,color,amssymb,amsmath,amscd}
\usepackage[normalem]{ulem}
\usepackage{cite}
\usepackage[]{graphicx}
\usepackage{makeidx}
\usepackage{amssymb}
\usepackage{epsfig,color}
\usepackage{pstricks,graphicx,epsfig,color,amssymb,amsmath,amscd}
\usepackage{cite}
\usepackage{amsmath}
\usepackage{nccmath}
\usepackage{setspace}
\usepackage{subcaption}
\usepackage{float}

\newcommand{\be}{\begin{eqnarray}}
\newcommand{\ee}{\end{eqnarray}}

\usepackage[]{caption}
\renewcommand\rho{\varrho}

\numberwithin{equation}{section}
\begin{document}

\begin{titlepage}
\title{Conversion of high frequency relic gravitational waves into photons in cosmological magnetic field}
\author{L. A. Panasenko$^{a}$, A. O. Chetverikov$^{a,b}$}

\maketitle
\begin{center}
$^a${Department of Physics, Novosibirsk State University, \\Pirogova 2, Novosibirsk 630090, Russia}\\
$^b${Voevodsky Institute of Chemical Kinetics and Combustion, Siberian Branch of the Russian Academy of Science, 3, Institutskaya St., Novosibirsk 630090, Russia}\\


\end{center}

\begin{abstract}

This work continues the research presented in the article \cite{Conversion}, where we estimate the Gertsenshtein effect's influence on the long-wavelength part of relic gravitational wave spectrum. Here, the differential equation system for the Gertsenshtein effect in Friedman–
LeMaitre–Robertson–Walker universe, derived in \cite{Conversion}, is simplified for gravitational waves in the under-horizon regime during radiation dominance epoch. Then, the obtained system is solved analytically. As a result of the solution analysis a conclusion was made about a significant increase of relic GWs with the frequencies $ k\gtrsim 10^{-11}$ Hz for magnetic field strength $B_0\sim 1$ nGs. In addition, at the end of the article model dependency of the result is discussed.

\end{abstract}
\thispagestyle{empty}
\end{titlepage}

\section{Introduction \label{s-intro}}
In the article \cite{Conversion} the equation system was derived for the conversion of gravitational wave (GW) into electromagnetic wave (EMW) in the presence of external magnetic field in curved space-time. After that, influence of that phenomenon on relic gravitational wave amplitude was estimated for low frequencies ($10^{-18}-10^{-16}$) Hz and the present day cosmological magnetic field amplitude $1$ nGs. 

In order to make the study \cite{Conversion} complete and consistent, it is important to check whether the influence of the Gertsenshtein effect on high-frequency gravitational waves is negligible. Indeed, if there is a significant enhancement or suppression of relic gravitational waves with high frequencies, we may have to change the predictions about their detectability in the present Universe for different inflation models.

In the previous work we divided wave vector $\boldsymbol{k}$ into two components: $\boldsymbol{k}||\boldsymbol{B}$ and $\boldsymbol{k}\bot\boldsymbol{B}$, where $\boldsymbol{B}$ is a cosmological magnetic field vector. Connection between metric perturbation and electromagnetic wave takes place only for perpendicular part of the wave vector. Moreover, separated systems were obtained for $h_{\times}$- and $h_{+}$-polarizations of metric perturbation tensor. In the manuscript we will focus on the first polarization.

A short discussion of results for $h_+$-polarisation is presented in Appendix A. The main conclusion is that there is no influence of the Gertsenshtein effect on gravitational wave amplitude for $h_+$-polarisation at all. But there may be generation of electromagnetic wave potential $f^y$ and generation of scalar metric perturbations $\Phi$ and $\Psi$.

All in all we conclude, that the Gertsenshtein effect changes the amplitude of gravitational wave only for the $h_{\times}$-polarization of the component propagating perpendicular to magnetic field $\boldsymbol{B}$. This is a well known result from works, where the Gertsenshtein effect was considered in the flat static space-time. For example non-cosmological problems are considered in \cite{g-to-gam1,g-to-gam2,g-to-gam3,g-to-gam4,g-to-gam5, g-to-gam6, g-to-gam7, g-to-gam10} and simple evaluations for cosmology and astrophysics are made in \cite{g-to-gam8, g-to-gam9}. 

To estimate the influence of the Gertsenshtein effect on relic gravitational waves with high frequencies let us, firstly, remind the previous results. In FLRW metric in the case $\boldsymbol{k}\bot\boldsymbol{B}$ and with taking into account interaction with a primary plasma during RD era from eq.(207) in  \cite{Conversion} we have:
 \begin{align}
	\label{SDEtoSolve}
	&f^x: a^2H^2{f}^{x \prime\prime}+aH^2\left[1+a\frac{H^{\prime}}{H}+8\frac{2B_0^2C_0-a^4}{16B_0^2C_0-a^4}+aH\Gamma\right]f^{x \prime}+\nonumber\\
	&+\left[\frac{k^2}{a^2}+2aHH^{\prime}-8H^2\frac{4B_0^2C_0+a^4}{16B_0^2C_0-a^4}+2\Gamma H+\omega_{pl}^2\right]f^x{-\gamma\beta H^2\left(af^{x\prime}+2f^x\right)}=-\frac{ikB_0}{a^4 m_{pl}}h_{\times},\nonumber\\
	&h_x^y: a^2 H^2 {h}_{\times}^{\prime\prime}+\left(4aH^2+a^2HH^{\prime}\right){h}_{\times}^{\prime}+\left[\frac{k^2}{a^2}+ \frac{16\pi G B_0^2}{{a^4}}\left(1-\frac{4B_0^2 C_0}{a^4}\right)\right]h_{\times} =\nonumber\\
	& = - \frac{16\pi GB_0 ik}{{a^2}}\left(1-\frac{16B_0^2C_0}{a^4}\right)m_{pl}f^x,
\end{align}
where prime means derivative over scale factor $a$. Denotations are the following: $h^x_y=h_{\times}$ -- the \{x,y\} component of the metric perturbation tensor, which is also the $h_{\times}$-polarization  for the case when initial gravitational wave propagates along $z$ axis; $f^x$ is an $x$-component of electromagnetic wave potential; $H=\frac{\dot{a}}{a}$ -- Hubble parameter; $B_0$ -- magnitude of cosmological (intergalactic) magnetic field in the present Universe\footnote{Physical magnetic field depends on the scale factor as $B(t)=\frac{B_0}{a(t)^2}$. So in the past the magnitude was higher. We have taken this fact into account when obtaining the system of Eq.(\ref{SDEtoSolve})}; $G=\frac{1}{m_{pl}^2}$ -- gravitational constant; $m_{pl}$ -- Planck mass; $k$ -- wave number; $C_0=\frac{\alpha^2}{90 m_e^4}$ -- constant in the Heisenberg-Euler action (see \cite{Conversion}), where $m_e$ -- electron mass, $\alpha$ -- fine structure constant. We assume that magnetic field is homogeneous\footnote{Today the commonly accepted model is a model of stochastic cosmological magnetic field. But it can be shown that for the simplest model, where the length of magnetic field coherence changes as $1/a$, gravitational wave front can not overcome the coherence length due to Universe expansion during radiation dominance era. So the approximation of homogeneous magnetic field allow to obtain some qualitative results and the top estimation of the Gertsenshtein effect influence on relic GWs} and directed along $x$ axis and $B_0=1$ nGs.

It is important to note, that previously we have taken into account the first loop-correction to the Maxwell action.
This correction is proportional to the product $CB^2$ and we accept, that it plays a significant role under the followong condition: $CB^2\gtrsim 10^{-5}$. This gives the condition for the magnetic field strength: $B \gtrsim 10^{10}$ Gs. The approximation of the effective Heisenberg-Euler action works for $B \ll m_e^2 \sim 10^{13}$ Gs. Therefore we ultimately accept the following limits of applicability of the conversion effect estimates using the full system (\ref{SDEtoSolve}):
\be\label{B-limits}
10^{10} \text{ Gs} \leqq B \leqq 10^{13} \text{ Gs}.
\ee
For a field strength less than $10^{10}$ Gs the system (\ref{SDEtoSolve}) is still be correct, but calculating the terms from the loop-correction is redundant.

We solve the equations for $10^{-9} \leq a \leq 10^{-4}$. In the chosen interval of the scale factor, the homogeneous magnetic field strength relative to the present value will be amplified by $10^8-10^{18}$ times. Thus, for the upper estimate of the effect on the amplitude of relic GWs, $B_0=1$ nG, the maximum value of the magnetic field strength is  $10^9$ Gs. Thus, the effect of the creation of a virtual electron-positron pair for the uniform field model can be neglected over the entire solution interval (see Eq. (\ref{B-limits})).

Thus, the Eq. (\ref{SDEtoSolve}) can be written in terms of time in the following form:
\be\label{eqhFinal01}
\begin{cases}
	&\left[\partial_t^2 + 3H \partial_t +\left(\frac{k^2}{a^2}-\frac{8\pi G B_0^2}{a^4}\right)\right]h_{\times}=-\frac{ik16\pi GB_0}{a^2} f^x,\\
	&\left[\partial_t^2 + 3H \partial_t +\frac{k^2}{a^2}\right]f^x=-\frac{ikB_0}{2a^4} h_{\times}.
\end{cases}
\ee
where we neglect all the terms, originated from the Heisenberg-Euler Lagrangian, and for a while omit the terms responsible for the interaction with the plasma (we will consider them further).

The article has the following structure. In the next section the system of differential equations (\ref{eqhFinal01}) is simplifyed for the high frequency limit, is made dimensionless and, finally, reduced to one differential equation of the forth order. Sec.\ref{ApproximateSol} is devoted to the analytical solution of the system. In Sec.\ref{Analysis} we analyze the solution and evaluate the Gertsenshtein effect influence on relic GWs amplitudes for different frequencies and magnetic field strengths. Finally, in Sec.\ref{CoherenceLenght}, \ref{Plasma} we discuss the impact of the magnetic field coherence length and introduce interaction with the primary plasma respectively. In conclusion we resume the work.

\section{Simplification of the system \label{Simplification}}
We can neglect the term  $\frac{8\pi G B_0^2}{a^4}$ under the following condition:
\be
k a \gg \sqrt{8 \pi} \frac{B_0}{m_{pl}}\approx 6,1 \times 10^{-23}\,\, \text{Hz}.
\ee
In order evaluate the effect influence we need to specify the problem. Further we will consider radiation domimamce (RD) epoch from the so called hadron epoch: the scale factor $a_1=10^{-9}$. Thus, we suggest that cosmological magnetic field was generated during the QCD phase transition\footnote{The QCD phase transition in the standart model of particles is a crossover, but there are hypothetical models, where it is a first-order phase transition and cosmological magnetic fields can be generated}.

Value of the multiplication $ka$ is minimal at $a_1$. Hence the condition of negligibility of the second term in round brackets on the left side of the first line of Eq. (\ref{eqhFinal01}):
\be\label{condition001}
k \gg 6,1 \times 10^{-14}\,\, \text{Hz}.
\ee

For values $a>a_1$ the right-hand side of the Eq.(\ref{condition001}) only increases, so the condition works on the entire solution interval.

Now let us make the replacement
\be
&\tilde{h}\equiv a(\eta)h_{\times}(\eta),\nonumber\\
&\tilde{f}\equiv a(\eta)f^x(\eta)
\ee
and to rewrite the Eq.() in terms of conformal time $\eta$. We obtain
\be
&\tilde{h}^{\prime\prime}-\frac{a^{\prime\prime}}{a}\tilde{h}+k^2\tilde{h}= -ik16\pi GB_0 \tilde{f}, \nonumber\\
&\tilde{f}^{\prime\prime}-\frac{a^{\prime\prime}}{a}\tilde{f}+k^2\tilde{f}= -\frac{ik\pi B_0}{2a(\eta)^2} \tilde{h}.
\ee
Here and below, the prime denotes the derivative with respect to $\eta$.

In the manuscript by high frequencies we mean frequencies satisfied the condition $k~\gg~a^{\prime} / a$. Gravitational waves with such frequencies are under-horizon mode for the considered interval of scale factor values: we can neglect the term proportional to $\frac{a^{\prime\prime}}{a}$. As a result, the final system has a very simple form of two coupled pendulums with a variable inhomogeneous part. Indeed, we have
\be\label{analyticalSystem}
&\tilde{h}^{\prime\prime}+k^2\tilde{h}= -\frac{ik16\pi B_0 }{m_{pl} \hbar} \tilde{f}, \nonumber\\
&\tilde{f}^{\prime\prime}+k^2\tilde{f}= -\frac{ik2\pi B_0 \tau_0^2}{m_{pl}\hbar\,\eta^2} \tilde{h}.
\ee
Here we replace dimentional electromagnetic potential by dimentionless $\tilde{\tilde{f^x}}\equiv \tilde{f^x}/m_{pl}$, immediately omitting the second tilde. Also we use that for RD epoch from $ad\eta=dt$ it follows that $\eta=2\sqrt{\tau_0 t}=2\tau_0 a$, where $\tau_0 = 35 \tau$  is a devider in the scale factor exprassion $a=\sqrt{\frac{t}{\tau_0}}$, $\tau$ -- the Universe lifetime. The frequency $k$ is substituted in hertz, therefore, for the accuracy of the dimensional matching, the Planck constant was returned.


In fact, for the RD epoch the term $\frac{a''}{a}=0$ during all the solution interval. That means that the system of Eqs.(\ref{analyticalSystem}) is correct for lower frequencies. Thus, the only condition we have to satisfy is the condition of Eq.(\ref{condition001})

Note also that the presence of an imaginary unit in the right-hand sides of the equations (\ref{analyticalSystem}) means that the phases of gravitational and electromagnetic waves are shifted relative to each other by $\pi/2$. This is expected since at the initial moment of time only the gravitational wave is present, and the amplitude of the electromagnetic wave is zero.

Let's discuss the solution plan. First, we need to obtain a fourth-order equation for one of the functions $\tilde{f^x}(\eta), \tilde{h}(\eta)$ from the equation system (\ref{analyticalSystem}). After that, we look for a solution in the form of an integral function of a complex variable. At the final stage of solution it is important to choose an optimal integration contour, which gives the simplest form of the solution.

The reader can immediately see that the solution will be a sinusoid with some modulating function.

We will derive the unknown coefficients of the solutions from the initial conditions. For arbitrary initial values of $h(\eta_1)\equiv h_1$ and $h'(\eta_1)\equiv h'_1$ we have
\be\label{InitConds}
\tilde{h}(\eta_1)&=&a_1 h_1,\nonumber\\
\tilde{h}^{\prime}(\eta_1)&=&\frac{h_1}{2\tau_0}+a_1 h'_1,\nonumber\\
\tilde{f}(\eta_1)&=&0,\nonumber\\
\tilde{f}^{\prime}(\eta_1)&=&0.
\ee
So we have assumed that electromagnetic wave is absent at the initial time. It is nessessary to explaine, why there can exist non-zero first derivative $h'_1$ at $\eta_1$: different frequencies enter under horizon at different times\footnote{Despite the fact that the term $a''/a$ is equal to zero for RD era, here we assume that for some stage the second term on the right side of Eqs.(\ref{analyticalSystem}) is negligible for beyond-horizon mode, therefore the GW amplitude is constant}. After the entry the dependency of GW amplitude roughly changes from constant (beyond-horizon mode) to $\tilde{h}(\eta)=\tilde{h}_0\frac{a(\eta_{entr})}{a(\eta)}cos(k\eta+\phi_{entr})$ \cite{SW}. Here $h_0$ -- relic GW amplitude at the end of inflation and $\phi_{entr}$ -- the phase, calculated from matching with constant mode at entry conformal time $\eta_{entr}$:
\be
\tilde{h}_0 cos(k\eta_{entr}+\phi_{entr})=\tilde{h}_0.
\ee
As a result, for different $k$, at the point $\eta_1$ we can obtain different phases $(k\eta_{1}+\phi_{entr})$ in cosine in $\tilde{h}_1=\tilde{h}_0\, cos(k\eta_{1}+\phi_{entr})$ and sine in $\tilde{h}'_1=-\tilde{h}_0 k \,sin(k\eta_{1}+\phi_{entr})$. The approximate time of entry can be calculated from the condition 
\be
k\eta_{entr}\sim 1.
\ee

As will become clear below, the frequency dependence does not change when the Gertsenshtein effect is taken into account. So for simplicity of presentation, we do not write these constant phases explicitly and present the result in terms of fraction of the amplitude obtained in the conventional theory of the tensor perturbations evolution.


To simplify further calculations, we introduce constants
\be
&C_1\equiv -\frac{ik16\pi B_0}{m_{pl}}\nonumber,\\
&C_2\equiv -\frac{ik2\pi B_0 \tau_0^2}{m_{pl}}.
\ee
Then, the system of equations (\ref{analyticalSystem}) takes the form
\be\label{analyticalSystem2}
&\tilde{h}^{\prime\prime}+k^2\tilde{h}= C_1 \tilde{f}, \nonumber\\
&\tilde{f}^{\prime\prime}+k^2\tilde{f}= \frac{C_2}{\eta^2} \tilde{h}.
\ee
The first constant is dimensional, the second -- dimensionless. So the right-hand sides of the equations, like all the terms on the left-hand side, have the dimension $\text{sec}^{-2}$.

Now we differentiate the first equation of the system (\ref{analyticalSystem2}) twice and obtain
\be
\tilde{h}^{\prime\prime\prime\prime}+2k^2 \tilde{h}^{\prime\prime}+k^4\tilde{h}=\frac{C_1 C_2}{\eta^2} \tilde{h}.
\ee

It will also be convenient to non-dimensionalize the function argument. Let us introduce
\be\label{xDef}
x\equiv \sqrt{C_1C_2}\, \eta \approx 3 i \times 10^{-3}\,k\,\eta.
\ee
Replacing all derivatives with respect to $\eta$ with derivatives with respect to $x$, we obtain a fairly simple expression
\be\label{ASEq}
\frac{d^4\tilde{h}}{dx^4}+2\gamma \frac{d^2\tilde{h}}{dx^2}+\gamma^2\tilde{h}=\frac{\tilde{h}}{x^2},
\ee
where we defined
\be\label{alpha007}
\gamma \equiv \frac{k^2}{C_1 C_2} = -\frac{m_{pl}^2}{32\pi^2 B_0^2 \tau_0^2} \approx -0.9 \times 10^5.
\ee
We emphasize that $\gamma$ does not depend on frequency $k$ and is less than zero.
\section{Analytical solution \label{ApproximateSol}}
In this section an approach is discussed to determining the analytical solution to Eq.(\ref{ASEq}) by integral representation. 

Considering Eq.(\ref{xDef}) and Eq.(\ref{alpha007}), let us redefine the variables $x = |x|$ and $\gamma = |\gamma|$. Then Eq.(\ref{ASEq}) becomes:
\be\label{eqSimle2}
\frac{d^4\tilde{h}}{dx^4}+2\gamma \frac{d^2\tilde{h}}{dx^2}+\left(\gamma^2+\frac{1}{x^2}\right)\tilde{h}=0.
\ee

Let us find a solution of this equation in the following form:
\be\label{funkForm}
\tilde{h}(x)=\int_{c} Z(y) e^{x y} dy,
\ee
where $c$ under the integral means an integration contour, along which the integrand is holomorphic. $Z(y)$ --  a function of a complex variable $y$.

By substituting Eq.(\ref{funkForm}) into Eq.(\ref{eqSimle2}) we obtain
\be
&\int_{c} Z(y)\left[F(y)x^2+1\right]e^{x y} dy = 0, \nonumber\\
&F(y)=\left(y^2+\gamma\right)^2.
\ee
It is necessary to exclude dependence on $x$. To do it, we integrate by parts twice
\be\label{eqWithoutBoardCond}
\int_{c}\left(\frac{d^2(ZF)}{dy^2}+Z\right)e^{x y}dy+xZFe^{x y}|_{\partial c}-\frac{d(ZF)}{dy}e^{x y}|_{\partial c}=0.
\ee
Based on the initial assumption that neither the integration contour nor the function $Z(y)$ depends on $x$, the only way to eliminate the last two terms in Eq.(\ref{eqWithoutBoardCond}) is to set to zero each of them. One can satisfy this condition by choosing an appropriate contour.

Finally, we have an equation for $Z(y)$:
\be\label{ZF}
\frac{d^2(ZF)}{dy^2}+Z=0.
\ee
and two solutions are:

\be\label{Z}
Z_{1,2}(y)=A_{1,2}\left(y+i\sqrt{\gamma}\right)^{-\frac{3}{2}\pm\frac{\beta}{2}}\left(y-i\sqrt{\gamma}\right)^{-\frac{3}{2}\mp\frac{\beta}{2}},
\ee
where
\be\label{beta}
\beta=\sqrt{1+\frac{1}{\gamma}},
\ee
$A_1$ and $A_2$ -- arbitrary constants. This function is meromorphic on the complex plane with a  branch cut between the points $\pm i\sqrt{\gamma}$.

Now we need to choose an appropriate integration contour to set two integration constant in Eq.(\ref{eqWithoutBoardCond}) to zero.

The first option is a closed contour. However, if this contour does not conclude singularity points $y=\pm i\sqrt{\gamma}$ of $Z_{1,2}(y)$, then $\tilde{h}\equiv 0$, which is a trivial solution to Eq.(\ref{eqSimle2}).

\begin{figure}[H]
\centering
\includegraphics[width=.7\linewidth]{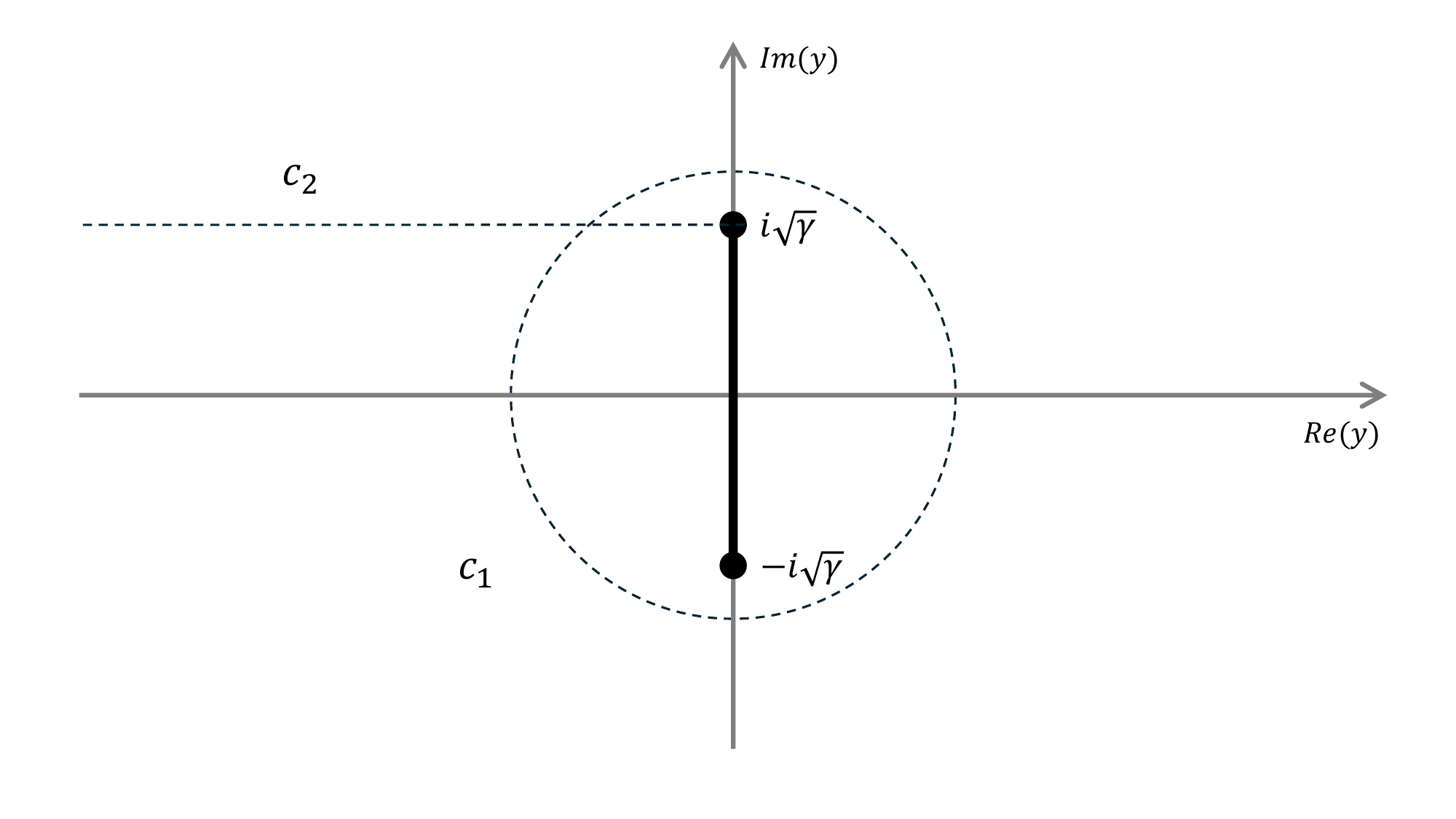}
\caption{\centering Two contours $c_1, c_2$ for solution in the integral form Eq.(\ref{funkForm}), which satisfy the conditions of the integration constants vanishing in Eq.(\ref{eqWithoutBoardCond}) and give linearly independent solutions. Bold line shows the branch cut.}
\label{fig:fig2}
\end{figure}

The second option is a closed contour enclosing an area containing one or both of these points. In this case, the integral equals the sum of residues at these points. Figure \ref{fig:fig2} shows an example of such a contour $c_1$. Then for the function $\tilde{h}$, up to an arbitrary constant, we have:
{\small
\be\label{Inth1}
\tilde{h}_{\pm}(x)=\oint_{c_1} \left(y+i\sqrt{\gamma}\right)^{-\frac{3}{2}\pm\frac{\beta}{2}}\left(y-i\sqrt{\gamma}\right)^{-\frac{3}{2}\mp\frac{\beta}{2}}e^{xy}dy.
\ee}


To calculate the integral in Eq.(\ref{Inth1}), we expand the integrand into a Laurent series:
\begin{gather}\label{LauSer}
\left(y+i\sqrt{\gamma}\right)^{-\frac{3}{2}\pm \frac{\beta}{2}}\left(y-i\sqrt{\gamma}\right)^{-\frac{3}{2}\mp \frac{\beta}{2}}e^{xy}= y^{-3}\sum_{k=0}^{\infty} c_k^\pm \left( \frac{i\sqrt\gamma}{y} \right)^k\sum_{m=0}^{\infty} \frac{\left( xy \right)^m}{m!}, \nonumber\\
c_k^\pm=\frac{\Gamma{\left(\pm\frac{\beta}{2}-\frac{1}{2}\right)}\ce{_2 F}_{1}\left(\pm\frac{\beta}{2}+\frac{3}{2},\,-k;\,\pm\frac{\beta}{2}-k-\frac{1}{2};\,-1\right)}{\Gamma{\left(\pm\frac{\beta}{2}-k-\frac{1}{2}\right)}k!},
\end{gather}
where $\Gamma$ is the Gamma function and $F$ -- the Hypergeometric function. Both series are convergent, since $|y|>\sqrt{\gamma}$ on the integration contour. 

Next, we replace the variables by $y=re^{i\phi}$, where $\phi$ runs from 0 to $2\pi$, $r$ is a constant radius  ($r>\sqrt{\gamma}$). Then for Eq.(\ref{Inth1}) we obtain:

\be\label{hpm}
\tilde{h}_\pm(x)=\int_0^{2\pi}\frac{e^{-2i\phi}}{r^2}\sum^{\infty}_{k=0}c^{\pm}_k e^{-ik\phi}\left(\frac{i\sqrt{\gamma}}{r}\right)^k\sum^{\infty}_{m=0}\frac{\left( xr \right)^m}{m!}e^{im\phi}d\phi=2\pi x^2\sum^{\infty}_{n=0}\frac{c_n^{\pm}\left(i\sqrt{\gamma}\,x\right)^k}{\left(n+2\right)!}.
\ee
In the resulting series, the terms can be regrouped in such a way as to isolate the exponent and obtain solutions in the form:
\be\label{h12}
\tilde{h}_{1,2}(x)=B_{1,2}e^{\pm i \sqrt{\gamma}x}x^2\,\, \ce{_1 F}_{1}\left(\frac{3}{2}+\frac{\beta}{2},3,\mp 2 i \sqrt{\gamma}\,x\right),
\ee
where $B_{1,2}$ are arbitrary constants.

Expression (\ref{h12}) can be obtained from Eq.(\ref{Inth1}) directly if we shift to one of the function branch points, go to integer powers and bypass this point on all sheets of the Riemann surface. Such a procedure is much more complicated from the point of view of mathematical justification, but immediately provides an answer in the form of Eq.(\ref{h12}).

Two more linearly independent solutions can be obtained by taking as the integration contour in Eq.(\ref{funkForm}) the straight line from the point $y_1=i\sqrt{\gamma}$ to $y_2=-\infty+i\sqrt{\gamma}$ ($c_2$ on the Figure \ref{fig:fig2}). At these points, the terms $Z_2 F e^{xy}$ and $\frac{dZ_2 F}{dy}e^{xy}$ turn to zero, because the degree of the monomial $\left(y-i\gamma\right)$ remains positive, even after taking the derivative, since $\beta>1$.

As a result, for the desired function we have:
\be
&\tilde{h}(x)=\int_{i\sqrt{\gamma}}^{-\infty+i\sqrt{\gamma}}\left(y+i\sqrt{\gamma}\right)^{-\frac{3}{2}-\frac{\beta}{2}}\left(y-i\sqrt{\gamma}\right)^{-\frac{3}{2}+\frac{\beta}{2}}e^{xy}dy=\nonumber\\
&=e^{i\sqrt{\gamma}x}\int_0^{-\infty} \left(y+2i\sqrt{\gamma}\right)^{-\frac{3}{2}-\frac{\beta}{2}}y^{-\frac{3}{2}+\frac{\beta}{2}}e^{xy}dy.
\ee
This integral, by differentiation by parts, reduces to the integral representation of the Tricomi function $U\left(\frac{3}{2}+\frac{\beta}{2},\,3,\, -2i\sqrt{\gamma}x\right)$. This function is the second linearly independent solution of Kummer's equation. Thus:
\be\label{h34}
\tilde{h}_{3,4}(x)=B_{3,4}e^{\pm i\sqrt{\gamma}x}x^2U\left(\frac{3}{2}+\frac{\beta}{2},\,3,\,\mp 2i\sqrt{\gamma}x\right).
\ee

It is worth noting that reducing the original Eq.(\ref{eqSimle2}) to Kummer's equation, using substitution $\tilde{h}(x)=e^{\pm\sqrt{\gamma}x}x^2 H(x)$, where $H(x)$ -- unknown function, is not a trivial task. At the same time, the approach described above with searching for solutions in integral form made it easy to find all the solutions.

\section{Analysis of the solution\label{Analysis}}
Let us analyze the obtained solutions Eqs.(\ref{h12}, \ref{h34}). The behavior of the amplitudes of these two pairs of solutions is determined by the following asymptotes:
\be
&\tilde{h}_{1,2}(x)\propto x^{\frac{1}{2}+\frac{\beta}{2}},\nonumber\\
&\tilde{h}_{3,4}(x)\propto x^{\frac{1}{2}-\frac{\beta}{2}}.
\ee
The first group gives rising functions, the second -- dropping down functions, because $\beta>1$ by definition in Eq.(\ref{beta}).

When $\beta$ is an odd integer, the Kummer and Tricomi functions reduce to simpler forms. For example, when $\gamma=1/8$ ($\beta=3$), the derived expressions allow the solution of the original equation to be written in the form:

\be\label{solOneEight}
&\tilde{h}_{\gamma=1/8}(x)= x^2e^{\frac{ix}{\sqrt{8}}}\left(B_1+B_2\Gamma\left(-2,\frac{ix}{\sqrt{2}}\right)\right)+x^2e^{\frac{-ix}{\sqrt{8}}}\left(B_3+B_4\Gamma\left(-2,\frac{-ix}{\sqrt{2}}\right)\right)
\ee

The final enhancement or attenuation on the finite interval $x$ depends on the initial conditions, while they determine the ratio between corresponding constants $B_i$. But it is important to note that $\gamma$ depends on magnetic field strength only. Hence there is a common degree of the growth for all the frequencies. For example the quadratically increasing solution Eq.(\ref{solOneEight}) corresponds to a magnetic field $B_0\approx 8.5\cdot 10^{-7}$ Gs. 

This strength is larger, than the upper limit on the present day value, which is obtained from observational data and theoretical predictions \cite{IMFlimits}. But we will see below that in the realistic models magnetic field strength decreases faster than $1/a^2$ during RD era \cite{IMF}. Therefore the initial $B_0=1$ nGs leads in the models to the higher strengths for considered interval $a\in\left[10^{-9},10^{-4}\right]$. Strictly speaking, for another model of magnetic field evolution we need to rewrite the system of Eqs.(\ref{SDEtoSolve}), but we can try to evaluate the qualitative behavior simply by increasing $B_0$. We will conduct a more rigorous analysis of specific models in a future study.

\subsection{Solution in the limit $\beta\rightarrow 1$}
The limit $\beta\rightarrow 1$ is equivalent to the condition $\gamma\rightarrow \infty$ or $\gamma^2 \gg \frac{1}{x^2}$. 
It is clear, that for some relations between solution interval $\left[x_1,x_2\right]$ and parameter $\gamma$ we can neglect the term $1/x^2$ in the Eq.(\ref{eqSimle2}). Indeed $\gamma$ does not depend on frequency and has a big absolute value (Eq.(\ref{alpha007})), while $\eta_1=2\tau_0 a_1=3.08 \cdot 10^{10}$ sec and, consequently from Eq.(\ref{xDef}) $x_1\approx 10^{8}k\,[\text{Hz}]$. As a result for the round brackets in Eq.(\ref{eqSimle2}) we have at the initial point $x_1$:
\be
\left(\gamma^2+\frac{1}{x_1^2}\right)\approx 0.8\cdot 10^{10}+\left(\frac{10^{-8}}{k\,[\text{Hz}]}\right)^2.
\ee

The general solution of the equation
\be\label{withoutxSq}
\frac{d^4\tilde{h}}{dx^4}+2\gamma \frac{d^2\tilde{h}}{dx^2}+\gamma^2\tilde{h}=0
\ee
has the form
\be\label{resonantTerms}
\tilde{h}=\left[A_1+A_2\,x\right] \cos(\sqrt{\gamma}\,x)+\left[A_3+A_4\,x\right] \sin(\sqrt{\gamma}\,x),
\ee
where $A_i$ are integration constants, which can be found from the initial conditions. 

We remind, that for free propagating GW the solution in expanding Universe is 
\be\label{hfree}
\tilde{h}_{free}\sim\cos(k \eta).
\ee
In contrast, in Eq.(\ref{resonantTerms}), we see resonant terms, proportional to $x\cos(k\eta)$. But the enhancement of the GW amplitude depends on the value of corresponding integration constants, that is, depends on the initial conditions.

Using initial conditions from Eq.(\ref{InitConds}) we obtain for the second and the third derivatives of $\tilde{h}$ at the initial point
\be\label{InitCond2}
\tilde{h}''(x_1)&=& -\frac{\tilde{f}(x_1)}{C_2}-\gamma \tilde{h}(x_1),\nonumber\\
h'''(x_1)&=& -\frac{\tilde{f'}(x_1)}{C_2}-\gamma \tilde{h'}(x_1),
\ee
where 
\be\label{InitCond3}
\tilde{f}(x_1)=0,\,\tilde{f}'(x_1)=0.
\ee
It is important to remind that we are working in variables $x=|x|$, $\gamma=|\gamma|$, starting with a Sec.\ref{ApproximateSol}.

Now we can find the constants $A_{i}$ explicitly, and the solution is
\be\label{resonant}
&\tilde{h}=\left(\tilde{h}(x_1)+\frac{\tilde{f}'(x_1)}{2\gamma C_2}(x-x_1) \right)\cos(\sqrt{\gamma}(x-x_1))+\nonumber\\
&+\left(\tilde{h}'(x_1)-\frac{\tilde{f}'(x_1)}{2\gamma C_2}-\frac{\tilde{f}(x_1)}{2\gamma C_2}(x-x_1) \right)\frac{\sin(\sqrt{\gamma}(x-x_1))}{\sqrt{\gamma}}.
\ee

Taking into account Eqs.(\ref{InitCond2}, \ref{InitCond3}) it can be rewritten as
\be\label{resonant2}
\tilde{h}=\tilde{h}(x_1)\cos(\sqrt{\gamma}\,(x-x_1))+\tilde{h}'(x_1)\frac{\sin(\sqrt{\gamma}(x-x_1))}{\sqrt{\gamma}}
\ee

There is no terms proportional to $x$ in the equation for the considered initial conditions.
Non-zero initial $\tilde{f}^x(x_1)$ gives higher values of $A_2$ and $A_4$. However, in the problem under consideration, the initial electromagnetic wave is absent. 

In the next section we show what is the difference between the result in approximation $\beta\rightarrow 1$ and the exact solution of Eq.(\ref{eqSimle2}). But here we stress that for very small magnetic field strengths there is no any change of the relic GW spectrum. That means, if precise measurements of intergalactic magnetic field give $B_0 \ll 1$ nGs, then the effect is absent.


At the end of the section let us, using conditions Eq.(\ref{InitConds}, \ref{InitCond2},\ref{InitCond3}), to compose a system for four integration constants in the exact anlytical solution Eq.(\ref{h12}, \ref{h34}) in a standard way:
\be
\begin{bmatrix}
&\tilde{h}& \\
&\tilde{h}^{\prime}& \\
&\tilde{h''}&\\
&\tilde{h'''}&
\end{bmatrix}(x_1)=
\begin{bmatrix}
\tilde{h}_1(x_1)& \tilde{h}_2(x_1)&\tilde{h}_3(x_1)&\tilde{h}_4(x_1)\\
\tilde{h}^{\prime}_1(x_1)& \tilde{h}^{\prime}_2(x_1)&\tilde{h}^{\prime}_3(x_1)&\tilde{h}^{\prime}_4(x_1) \\
\tilde{h}^{\prime\prime}_1(x_1)& \tilde{h}^{\prime\prime}_2(x_1)&\tilde{h}^{\prime\prime}_3(x_1)&\tilde{h}^{\prime\prime}_4(x_1)\\
\tilde{h}^{\prime\prime\prime}_1(x_1)& \tilde{h}^{\prime\prime\prime}_2(x_1)&\tilde{h}^{\prime\prime\prime}_3(x_1)&\tilde{h}^{\prime\prime\prime}_4(x_1)
\end{bmatrix}
\begin{bmatrix}
B_1 \\
B_2 \\
B_3\\
B_4
\end{bmatrix},
\ee
where $\tilde{h}_i$ are defined in Eqs.(\ref{h12}, \ref{h34}).

\subsection{Results of the Cauchy problem solution}
\begin{figure}[H]
\begin{subfigure}{.5\textwidth}
  \centering
  \includegraphics[width=.7\linewidth]{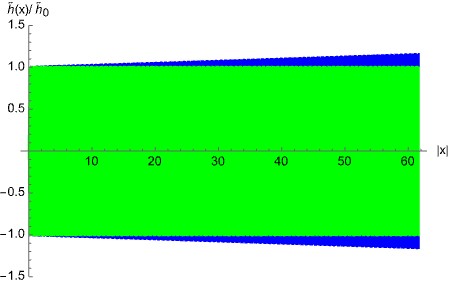}
  \caption{$k=5\cdot 10^{-12}$ Hz}
  \label{fig:sfig1}
\end{subfigure}%
\begin{subfigure}{.5\textwidth}
  \centering
  \includegraphics[width=.7\linewidth]{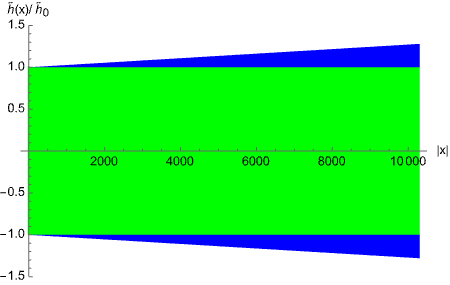}
  \caption{$k=10^{-9}$ Hz}
  \label{fig:sfig2}
\end{subfigure}
\caption{\centering GW amplitude dependence on $x=3\cdot 10^{-3} k\eta(a)$ at the interval $a\in \left[10^{-9}, 10^{-4}\right]$ for $B_0=1$ nGs: blue -- analytical solution of Eq.(\ref{eqSimle2}) according to Eqs.(\ref{h12},\ref{h34}); green -- analytical solution for the free propagating gravitational wave. Amplitude is normalized by the initial GW amplitude $\tilde{h}_0$. High frequency oscillations are not distinguishable} 
\label{fig:fig1}
\end{figure}

The Figure \ref{fig:fig1} shows the plots of analytical solutions of Eq.(\ref{eqSimle2}) and, for comparison, of the analytical solution for free propagating GW $\tilde{h}_{free}=\tilde{h}_0 \cdot \cos{k\eta}=\tilde{h}_0 \cos{\sqrt{\gamma}x}$ and different parameters $k$. All the plots are normalized by the initial amplitude $\tilde{h}_0$.
Almost the same picture as on the right panel of Figure \ref{fig:fig1} is obtained for all the frequencies $k>1$ nHz:
 thirty percent amplification of the GWs amplitudes.

We, finally, have to consider $B_0$ as a free parameter of the task. Indeed, the exact magnetic field strength has not yet been measured. In addition, in realistic models magnetic field magnitude decreases faster then $1/a^2$ due to dissipation in the turbulent plasma. That means that maximum magnetic field magnitude and, consequently, a strength of the Gertsenshtein effect may be even higher for the considered interval $a \in \left[10^{-9},10^{-4}\right]$\footnote{The moment of initial magnetic field origin $a_1$ is also a model dependent parameter}. 

\begin{figure}[H]
\begin{subfigure}{.5\textwidth}
  \centering
  \includegraphics[width=.8\linewidth]{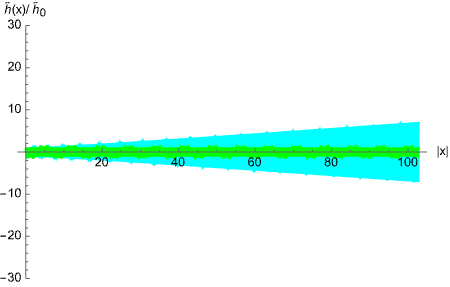}
  \caption{$k= 10^{-12}$ Hz}
  \label{fig:sfig1}
\end{subfigure}%
\begin{subfigure}{.5\textwidth}
  \centering
  \includegraphics[width=.8\linewidth]{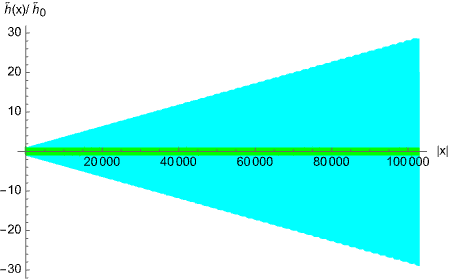}
  \caption{$k=10^{-9}$ Hz}
  \label{fig:sfig2}
\end{subfigure}
\caption{\centering GW amplitude dependence on $x=3\cdot 10^{-3} k\eta(a)$ at the interval $a\in \left[10^{-9}, 10^{-4}\right]$ for $B_0=0.1$ nGs: blue -- analytical solution of Eq.(\ref{eqSimle2}) according to Eqs.(\ref{h12},\ref{h34}); green -- analytical solution for the free propagating gravitational wave. Amplitude is normalized by the initial GW amplitude $\tilde{h}_0$. High frequency oscillations are not distinguishable} 
\label{fig:fig5}
\end{figure}

For example, for one of the models with the initial kinetic helicity of the plasma, the law of change of magnetic field magnitude during RD era is $B\propto 1/a^3$ \cite{IMF}. This increases the magnetic field strength by five orders of magnitude on average (for the considered interval).
Of course, for any model we must modify the system of Eqs.(\ref{SDEtoSolve}), and, where necessary, take into account the loop correction of light scattering on light. However, a naive estimate of the effect for a magnetic field, an order of magnitude greater than the initially accepted 1 nG, entails an increase in the amplitude of relic GWs by more than an order of magnitude for a frequency of $10^{-10}$ Hz (Figure \ref{fig:fig5}).
Here we obtain an increase of more than an order of magnitude for GWs with $k>10^{-12}$ Hz.

In the end of this section we can conclude that the Gertsenshtein effect increases the amplitudes of relic GWs for frequencies $k\gtrsim 10^{-11}$ Hz for the considered magnetic field magnitudes.



\section{Coherence length of the cosmological magnetic field\label{CoherenceLenght}}
So far we have made estimates of the GWs amplitude by the end of the RD era under the assumption of a homogeneous magnetic field. It is practically equivalent to the condition of a very large coherence length.

If we use the generally accepted model of a stochastic magnetic field, then, depending on its coherence length, we can get either the same effect or a much weaker one. This is due to the fact that when a gravitational wave overcomes a characteristic distance where the magnetic field strength can be considered constant, the direction of the magnetic field will randomly change. This will lead to a change in the direction of the driving force on the right side of Eqs. (\ref{analyticalSystem2}). As a result, the strengthening of the GW will be replaced by weakening, and on average, the effect of enhancement the amplitude over a large number of coherence lengths will become zero.

Let us estimate the coherence length of the primordial magnetic field (PMI) by the end of the RD stage, and find out whether the gravitational wave front manages to overcome at least one coherence length before the moment $a_2=10^{-4}$. It is important to understand that the coherence length in the above models grows quickly, and a wave front moving at the speed of light may not catch up with it.

We will make the estimates for the law of change of quantities (magnetic field strength and coherence length) only due to the expansion of the Universe, and will not take into account the energy losses due to heating of the primary plasma and its turbulence. 

The reader can see for example the article \cite{IMF}, where the authors obtain exact dependencies of the coherence length and magnetic field magnitude for different models of PMI generation by the first-order phase transitions\footnote{The initial coherence length in such models is comparable with the characteristic size of the bubbles of a new phase}. The general conclusion is that the comoving coherence length increases with time due to the magnetic field suppression at smaller scales. Hence, in the realistic models $\lambda_{coh}$ increases even faster than $a(t)$.

So, we have $\lambda_{coh} (a) = \lambda_{coh\,0} \,a$, where $\lambda_{coh\,0}$ -- coherence length of the present day intergalactic magnetic field. The wave front moves at the speed of light. Let us determine the value of the scale factor when the wave front crosses $\lambda_{coh} (a)$ for the RD era:
\be\label{evalCohL}
&c\left(t-t_1\right)=\lambda_{coh\,0} a,\\
&a(t)=\sqrt{\frac{t}{\tau_0}},
\ee
where $t_1$ is the initial time, $t>t_1$, $\tau_0 \approx 35 \tau$ and $\tau=4.4 \times 10^{17}$ sec is the lifetime of the Universe.
Let us rewrite the Eq.(\ref{evalCohL}) in terms of the scale factor:
\be
a\left[1-\left(\frac{a_1}{a}\right)^2\right]=\frac{\lambda_{coh\,0}}{c\tau_0}.
\ee
Substituting the value $a_1 = 10^{-9}$ and $\tau_0$ we get
\be
a\left[1-\left(\frac{10^{-9}}{a}\right)^2\right]\approx 6.8\times 10^{-6} \frac{\lambda_{coh\,0}}{1 \, \text{Mpc}}.
\ee
The motivation to use a characteristic scale 1 Mpc is that current observations of the photon Compton cooling are not sensitive to coherence lengths $\lambda_{coh\,0}>1$ Mpc.

\begin{figure}[H]
\centering
\includegraphics[width=.7\linewidth]{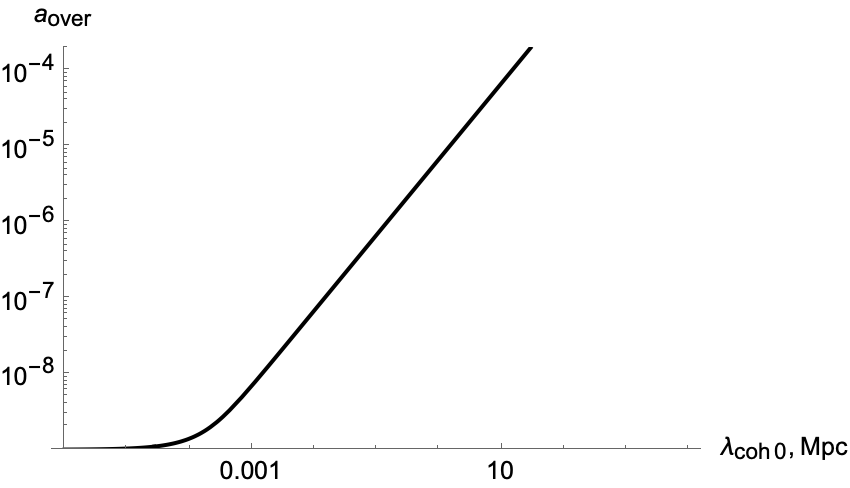}
\caption{\centering The first intersection scale factor for RD era as a function of the coherence length in the modern Universe} 
\label{fig:figRD}
\end{figure}

Let us substitute the scale factor of overcoming $a_{over}=x\cdot 10^{-9}$, where $1<x\leq 10^5$, and obtain a quadratic equation 
\be
x^2-A x -1 =0,
\ee
with the roots
\be
x_{1,2}=\frac{1}{2}\left(A\pm\sqrt{A^2+4}\right)
\ee
where $A=6.8\times 10^{3} \frac{\lambda_{coh\,0}}{1 \, \text{Mpc}}$. 
Positive $x$ can be obtained only for "$+$" sign. On the Fig.(\ref{fig:figRD}) the first intersection scale factor $a_{over}=x \cdot 10^{-9}$ for RD era is plotted as a function of $\lambda_{coh \,0}$.


It is important to note that the fact that for the value $\lambda_{coh\,0} \sim 10$ Mpc the GW begins to catch up with the coherence length by the end of the RD epoch does not mean that the effect is zero. Of course, it diminishes, but on average it is not equal to zero\footnote{We should remember that magnetic field magnitude also becomes smaller as $1/a(t)^2$ and the conversion in both directions ($g\leftrightarrow\gamma$) becomes weaker for later times of the Universe evolution}. At the same time, the smaller the coherence length of the magnetic field today, the negligibly small the change in the relic GW amplitude due to GWs conversion into electromagnetic waves.

It is worth to emphasize once again that the research objective is the relic GW spectrum
that was not so far from the end of RD era. Nevertheless it is an interesting question: does the Gertsenshtein effect affect on relic GWs during matter dominance (MD) era? 
We can similarly evaluate the overcoming scale factor. Indeed, let us replace the law of change of the scale factor with time by
\be
a= \left(\frac{t}{\tau}\right)^{2/3}
\ee
and substitute in Eq.(\ref{evalCohL}) and $a_{MD\,1} = 10^{-4}$ we get
\be\label{eqCondition}
a_{over}=10^{-4}\left(2.4\cdot 10^{-2}\frac{\lambda_{coh 0}}{1\,\text{Mpc}}-1\right)^2
\ee
By definition $a_{over}>a_{MD\,1}$. Hence
\be
\left(2.4\cdot 10^{-2}\frac{\lambda_{coh 0}}{1\,\text{Mpc}}-1\right)>1,
\ee
and solution of this equation exists for $\lambda_{coh \,0}\geq 80$ Mpc. For smaller values of $\lambda_{coh \,0}$ the gravitational wave front will not catch up with the coherence length of cosmological magnetic field.

\begin{figure}[H]
\centering
\includegraphics[width=.7\linewidth]{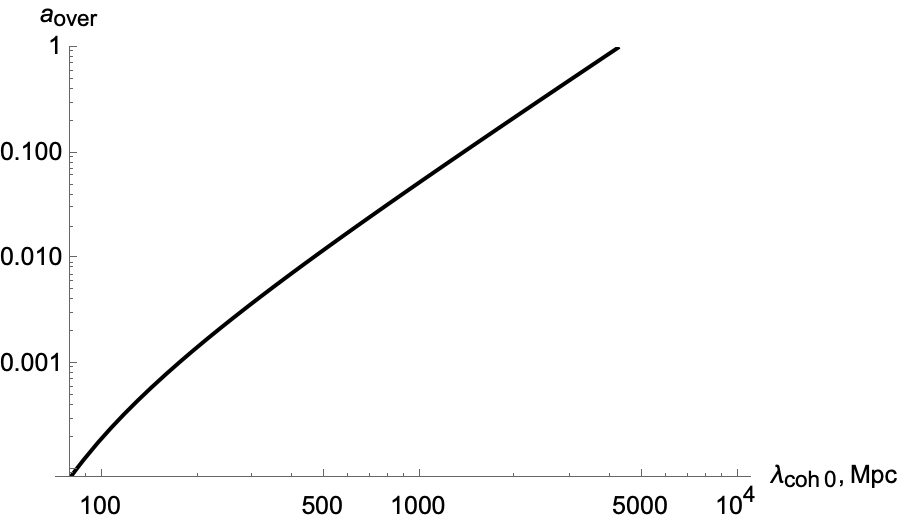}
\caption{\centering The first intersection scale factor for MD era as a function of the coherence length in the modern Universe} 
\label{fig:fig3}
\end{figure}

Fig.(\ref{fig:fig3}) shows the first intersection scale factor for MD era as a function of $\lambda_{coh \,0}$.

Depending on $\lambda_{coh \,0}$, after the GWs have passed the MD epoch, the spectral change accumulated during the RD epoch may be somehow neutralized. In this case, if the effect is large enough to be observable, it will manifest itself more strongly in the GW spectrum obtained from the CMB observations then in the spectrum from observations in the modern Universe\footnote{Space interferometers or pulsar timing array}.

However, even during MD era, the conversion of GW into EMW occurs, but under the influence of a magnetic field of smaller amplitude than during the RD era. Therefore, there may be two counteracting effects here.

Indeed, if the coherence length is too small, there will be no influence of the Gertsenshtein effect on relic GWs during the RD epoch. Then we can expect its manifestation during the MD epoch, due to the rapid growth of the coherence length.

To summarize this discussion, we can conclude that the coherence length evolution is model dependent and this model dependence affects the final strength of the Gertsenshtein effect influence on the relic GWs spectrum (whether it is a spectrum from recombination epoch or a modern spectrum). 
It is fair to note that for a coherence length of $\lambda_{coh\,0}\sim 10$ Mpc, the homogeneous magnetic field approximation is valid throughout both the RD and MD epochs. But ultimately, the crucial meaning in solving this issue has an accurate measurement of the parameter $\lambda_{coh\,0}$, which may become possible in the future\footnote{To date some limits on the magnetic field coherence length can be obtained from the observation of Compton cooling of photons, but this method is sensitive to coherence lengths $\lesssim 1$ Mpc.}.

\section{Interaction with primary plasma \label{Plasma}}
The last important effect that we must consider is the interaction of electromagnetic wave with the primary plasma during RD era. There are two parts of this effect, which are discussed in \cite{Conversion}: damping factor $\zeta$ and plasma frequency $\Omega_{pl}$. The equation for plasma frequency for relativistic particles with $m< T$ has the form
\be
\Omega^2_{pl}=\frac{2T^2}{9}\sum_j e_j^2,
\ee
where the summation is done over all relativistic charged particles with charges $e_j$.
So, we will evaluate $\Omega^2_{pl}\sim \alpha T^2$.

For the damping term we use the following assessment 
\be
\zeta =c\sigma n\sim \alpha^2 T,
\ee
where $c$ -- the seed of light, $n=0.1g_* T^3$ -- the density of charged particles in plasma, $g_* = 10 - 100$ is the number of charged particle species and $\sigma = \alpha^2/T^2$ -- the scattering cross-section.

We will substitute this terms as the functions of $a(\eta)$: 
\be
\Omega_{pl}^2(\eta)=\Omega_{pl}^2(a_2)\,\, \frac{a_2^2}{ a(\eta)^2},\qquad
\zeta(\eta)=\zeta(a_2) \frac{a_2}{ a(\eta)},
\ee
where $a_2=a(\eta_2)=10^{-4}$. Recalculating temperature of recombination $T(a=1/1100)=3000$ K, we obtain
\be\label{GammaNegl}
&\Omega_{pl}(a_2)\approx 5\cdot 10^9\, \text{Hz},\qquad\zeta(a_2)\approx 3\cdot 10^{6}\, \text{Hz}.
\ee
The second line in Eq.(\ref{analyticalSystem2}) takes the form
\be
\tilde{f}''+\left[k^2+\Omega^2_{pl}(a_2) a_1^2-\zeta(a_2)a_1 \frac{a'}{a}\right]\tilde{f}+\zeta(a_2) a_1 \tilde{f}'=\frac{C_2}{\eta^2}\tilde{h},
\ee
where all the derivatives are with respect to conformal time $\eta$. 

Finally, we obtain a fourth-order equation for $\tilde{f}$
\be\label{PlasmaInter}
&\tilde{f}''''+\zeta \tilde{f}'''+\left[2k^2-\Omega^2-\frac{\zeta}{\eta}\right]\tilde{f}''+\zeta\left[\frac{2}{\eta^2}+k^2\right]\tilde{f}'\nonumber\\
&+\left[k^2\left(k^2-\Omega^2-\frac{\zeta}{\eta}\right)-\frac{C_1C_2}{\eta^2}-\frac{2\zeta}{\eta^3}\right]\tilde{f}=0,
\ee
where we replaced $\zeta\equiv \zeta(a_2)a_1$ and $\Omega\equiv\Omega_{pl}(a_2)a_1$.

The Eq.(\ref{PlasmaInter}) is much more complicated than the Eq.(\ref{analyticalSystem2}) and its solution is beyond the scope of the article. Apparently, the interaction with the plasma can suppress the effect of the GW amplitude amplification obtained in the previous sections. However, in addition, there may be a characteristic imprint of this effect on the relic GW spectrum. 

We emphasize that without the Gertsenshtein effect interaction with the plasma would not occur, since the graviton itself is electrically neutral. In the problem, interaction with the plasma is due to the conversion of graviton into photon under the influence of the cosmological magnetic field.

We will devote our further study to the analysis of this problem.

\section{Conclusion \label{s-concl}}
In the present work we simplified for the high frequencies the system from \cite{Conversion}, describing conversion of relic GWs into EMWs under influence of the cosmological magnetic field in expanding Universe. We have found the exact analytical solution, analyze it and solve the Cauchy problem. 
In the last two sections the stochastic magnetic field model was discussed and the differential equation was obtained for taking into account interaction with primary plasma.

The conclusion was made, that there is a significant enhancement of relic GWs with the frequencies $k\gtrsim 10^{-11}$ Hz for the magnetic field strength $B_0\sim 1$ nGs. For $B_0 \ll 1$ nGs we showed that there is no amplification effect, but this result is model-dependent. 

Indeed, we have shown that it is impossible to exclude the model dependence of the result on the coherence length of the cosmological magnetic field. There remains hope for the results of measurements based on observations.

All in all, the distortion of the spectrum of relic GWs due to the considered effect strongly depends on the model of the magnetic field generation and evolution. On the other hand, this characteristic dependency may become a potential method for verifying these models.

The first planned stage of the future study is to analyze and solve the Eq.(\ref{PlasmaInter}), which takes into account the interaction of generated electromagnetic wave with the primary plasma during RD era.

The second planned stage is to estimate the effect on GW amplitudes caused by the Gertsenshtein effect during MD era. It seems, that the analytical approach used in the work is applicable to this problem too. 

Let us note also that the most interesting frequencies for MD epoch lye in the range $10^{-9}-10^{-3}$ Hz because of sensitivity of the the pulsar timing array and the sensitivity of the future space interferometers \cite{PTA, eLISA, DECIGO}. The considered in our work effect, along with other possible effects (see for example \cite{nHzGWs,nHzGWs2,nHzGWs3}), may contribute to the signal of nanohertz GWs, which PTA experiment apparently sees \cite{PTAsignal}.


\section*{Acknowledgement}
We thank A.D. Dolgov, A.P. Ulyanov and E.A. Lashina for the very important discussions.\\
The work was supported by RSF Grant 23-42-00066.

\section*{Appendix A}
Let us rewrite Eq.(24) and Eq.(154) from \cite{Conversion}:
\be\label{GWsys1}
	&\left[\partial_t^2+3H\partial_t-\frac{\Delta}{a^2}+3\left(\frac{\ddot{a}}{a}-4H^2\right)\right]h_{00}-\frac{1}{2}\partial^2h+\left(4H^2-\frac{\ddot{a}}{a}\right)h^i_i=-16\pi G T^{EM(1)}_{00},\nonumber\\
	&2H\left[\partial_j h_{00}+\frac{\partial_x h_{xj}+\partial_y h_{yj}+\partial_z h_{zj}}{a^2}\right]=-16\pi G T^{EM(1)}_{0j},\nonumber\\
	&\left[\partial_t^2+3H\partial_t-\frac{\Delta}{a^2}\right]h^i_{j}+\delta^i_{j}\left[-\frac{\partial^2 h}{2}+\left(\frac{\ddot{a}}{a}+2H^2\right)h_{00}+\frac{\ddot{a}}{a}h^l_l\right]=-16\pi G T^{i\,EM(1)}_{j},\,\,\,\,\,
\ee
\be\label{emwY}
	&\left[ \ddot{f_y} - \frac{\Delta f_y}{a^2} + H \dot{f_y} +B\partial_z h^{y}_y  \right]   = 0,
	\ee
    where dot means derivative with respect to time $t$, $T^{\mu\,EM(1)}_{\nu}$ -- correction to electromagnetic energy-momentum tensor (EMT) in the first perturbation order, $f_y=-a^2 f^y$ and $f^y$ are covariant and contravariant $y$-component of the electromagnetic wave potential and
\be 
h^i_i~\equiv~h^x_x~+~h^y_y~+~h^z_z~=~-\frac{h_{xx}+h_{yy}+h_{zz}}{a^2}.
\ee
EMT corrections are also calculated in \cite{Conversion} in Eqs.(134, 138-143). After simplification we have
\be\label{EMTBx0j}
	&T^{EM(1)}_{0j}=B\left[\dot{f}_{y}\delta_{jz}-\dot{f}_{z}\delta_{jy}\right],\nonumber\\
	&T^{x\,EM(1)}_x=\frac{1}{2}\left[2Bf^y_{.\,\,z}-B^2\left(h^y_y+h^z_z\right)\right],\nonumber\\
	&T^{x\,EM(1)}_y=-Bf^x_{.\,\,z}+B^2h^x_y,\nonumber\\
	&T^{x\,EM(1)}_z=Bf^x_{.\,\,y}+B^2h^x_z,\nonumber\\
	&T^{y\,EM(1)}_y=T^{z\,EM(1)}_z=-Bf^y_{.\,\,z}+B^2\frac{h^y_y+h^z_z}{2},\nonumber\\
	&T^{y\,EM(1)}_z=0,
\ee
where $\delta_{ij}$ -- Kronecker symbol, $f^i_{.\,j}$ -- Maxwell tensor for electromagnetic field.

It is important to remind that in this manuscript we neglect all the terms, originated from the Heisenberg-Euler action (proportional to $C B^2$, where $C=C_0$ for low temperatures), but in \cite{Conversion} we do not.

From Eqs.(\ref{GWsys1}, \ref{emwY}) and EMT corrections Eq.(\ref{EMTBx0j}) we obtain the system for $\{h_{+}, f^y, \Phi, \Psi\}$ in case $\boldsymbol{k} \bot \boldsymbol{B}$ ($\boldsymbol{k}||\boldsymbol{Oz}$):
\be
\begin{cases}\label{eqsys100}
	&\left[\partial_t^2 + 3H \partial_t +\left(\frac{k^2}{a^2}-\frac{8\pi G B_0^2}{a^4}\right)\right]h_+=\frac{16\pi GB_0}{a^2}\left(ikf^y+3\frac{B_0}{a^2}\Psi\right),\\
	&\left[\partial_t^2 + 3H \partial_t +\frac{k^2}{a^2}\right]f^y=\frac{ikB_0}{2a^4}\left[h_+-3\Phi+8\Psi\right],\\
	&ikH\left(\Phi+\Psi\right)=-\frac{4\pi G B_0}{a^2} \dot{f}^y,\\
	&\Psi = -\frac{ika^2}{3B_0}f^y.
\end{cases}
\ee
where we used  expansion of the metric perturbation tensor $h^{\mu}_{\nu}$ on scalar $\{\Phi,\Psi\}$ and traceless tensor $\{h_+,h_{\times}\}$ modes\footnote{Vector modes of metric perturbations are not significant in cosmology}\cite{SW}
\be
	&h_{tt}=2\Phi(t,\bf{r}),\nonumber\\
	&h^z_{z} = 2 \Psi(t,\bf{r}),\nonumber\\
	&h^x_x = 2 \Psi(t,\bf{r})+ h_{+}(t,\bf{r}),\nonumber\\
	&h^y_y = 2 \Psi(t,\bf{r})- h_+(t,\bf{r}),\nonumber\\
	&h^x_y= h_{\times}(t,\bf{r}).
\ee
To obtain the first equation in the system Eq.(\ref{eqsys100}), the equation for $h^y_y$ was subtracted from the equation for $h^x_x$ (see the system of equations (\ref{GWsys1})). The third equation was obtained for the component $h_{0z}$, and the fourth by adding the equations for $h^x_x$ and $h^y_y$ with simultaneous subtraction of the doubled equation for $h^z_z$. 

After substituting the fourth line of Eq.(\ref{eqsys100}) into the first line of the same system, we obtain an equation for the $h_+$ GW polarization in the case of $\boldsymbol{k} \bot \boldsymbol{B}$:
\be
\left[\partial_t^2 + 3H \partial_t +\left(\frac{k^2}{a^2}-\frac{8\pi G B_0^2}{a^4}\right)\right]h_+=0.
\ee

This result is very interesting and requires explanation. The tensor GW propagating in the magnetic field generates EMW, and the EMW in turn generates scalar metric perturbations. But the EMW and $\Psi$ are related in such a way that their corresponding components of the energy-momentum tensor completely compensate each other. As a result, the amplitude of the tensor GW $h_+$ remains unchanged (in the considered order of perturbation).

From the above, we conclude that the Gertsenshtein effect affects the amplitude of the gravitational wave only for the $h_{\times}$-polarization of the GW component, propagating perpendicular to $\boldsymbol{B}$.

The aim of the research is to investigate the Gertsenshtein effect influence on the amplitude of GWs, therefore we will leave the analysis of scalar and electromagnetic perturbations evolution outside the scope of the current study.



\color{black}

\end{document}